# Class Balanced PixelNet for Neurological Image Segmentation

Mobarakol Islam
NUS Graduate School for Integrative Sciences and Engineering (NGS), National University of Singapore, Singapore
mobarakol@u.nus.edu

Hongliang Ren
Department of Biomedical Engineering, National University of Singapore, Singapore
ren@nus.edu.sg

## ABSTRACT
In this paper, we propose an automatic brain tumor segmentation approach (e.g., PixelNet) using pixel level convolutional neural network (CNN). The model extracts feature from multiple convolutional layers and concatenates them to form a hyper-column where samples a modest number of pixels for optimization. Hyper-column ensures both local and global contextual information for pixel wise predictor. The model confirms the statistical efficiency by sampling few number of pixels in training phase where spatial redundancy limits the information learning among the neighboring pixels in conventional pixel-level semantic segmentation approaches. Besides, label skewness in training data leads the convolutional model often converge to the certain classes which is a common problem in the medical dataset. We deal this problem by selecting an equal number of pixels for all the classes in sampling time. The proposed model has achieved promising results in brain tumor and ischemic stroke lesion segmentation datasets.

## 1. INTRODUCTION

In medical image analysis, deep learning (DL) models such as the convolutional neural network (CNN) is providing exciting solutions and significant advances have been made to achieve state-of-the-art performance. Magnetic resonance imaging (MRI) is a widely-used imaging technique since it is possible to acquire complementary information from different modalities and sequences of images. As a 3D imaging technique, MRI is the most used medium in neuroimaging to diagnose and assess malformation of the brain such as brain tumor (gliomas) and stroke. It is very important to estimate the relative volume of the lesion as well as segment it to monitor the progression, plan the treatment and follow-up studies.

Recently, deep learning (DL) models have shown outstanding performance in medical applications such as classification, detection, and segmentation [27, 28]. DL model like convolutional neural networks (CNN) is capable of learning high level and task adaptive hierarchical features from MRI training data and take part as an effective segmentation approach in neuroimaging. Havae et al. [1] build a CNN based two-pathway cascade network which performs a two-phase training using both local and global contextual features and tackle difficulties related to imbalance of tumor labels in data. Another similar approach DeepMedic [2] use two convolutional parallel pathways and 3D CNN architecture with 11-layers for brain lesion segmentation. Later, a modified version of DeepMedic with residual connection utilize for brain tumor segmentation [3]. On the other hand, Pandian et al. [4] and Casamitjana et al. [5] use 3D volumetric CNN to train sub-volume of multi-modal MRIs and show that 3D CNN performs well for segmentation as MRI acquires 3D information. The benefit of these architectures is that they performed well with the comparatively smaller dataset. However, they are computationally expensive as it needs 3D kernels and enormous number of trainable parameters. Alex et al. [6] use 5 layers deep Stacked Denoising Auto-Encoder (SDAE) and Randhawa et al. [7] use 8-layer CNN and Pereira et al. [8] use deeper CNN architecture with a small kernel for segmenting gliomas from MRI. Fully Convolutional Network (FCN) which is a neural network composed of convolutional layers without any fully-connected layer, can be trained end-to-end, pixels-to-pixels, exceed the state of art in semantic segmentation. Chang et al. [9] utilize FCN with hyperlocal features applying bilinear interpolation for gliomas segmentation. An FCN and conditioner random field (CRF) have been applied for tumor segmentation with 3 modalities of MRI brain images [10]. However, all these architectures utilize local contextual information by using patch-wise learning where global information of training images is out of consideration.

On the other hand, deep learning models like convolutional neural networks are also utilizing to segment localize, extent and evolution of ischemic stroke lesion. Guerrero et al. [11] utilize u-shaped residual network (uResNet) to segment stroke lesion to differentiate with white matter hyper-intensity. A deep symmetry convnet with two symmetric patches feed into the network at the same time for stroke lesion segmentation [12]. Shen et al. [13] combine CNN and handcrafted feature for ischemic stroke lesion segmentation. Choi et al. [14] utilize volumetric 3D CNN for stroke lesion outcome prediction. However, all these approaches either less accurate or ignore some particular cases of stroke lesion like stroke lesions spread not only in one hemisphere but in the whole brain.

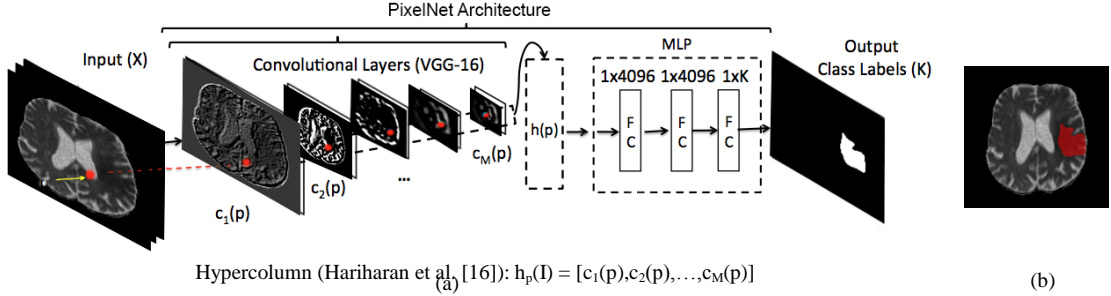

**Figure 1. (a) PixelNet architecture with stroke lesion segmentation example (b) An example of segmented mask with ADC modality.**

In pixel-wise segmentation with convolutional predictors, stochastic gradient descent (SGD) treats the training data independently and predicts every pixel separately [18]. Specially max-pooling and downsampling (striding) form spatial insensitivity in semantic segmentation which limits the spatial accuracy in deeper convolutional layers. To minimize this hurdle and ensure the contextual information in higher convolutional layers, many predictors have built with multiscale features. Hariharan et al. [16] build "Hypercolumns" by extracting features of the same pixels from multiple layers and form a vector. In [17], an FCN efficiently builds to extract features by linear prediction in a coarse to fine manner. Deeplab [18] incorporates filter dilation and linear-weighted fusion in fully connected layers to reduce memory footprint. ParseNet [19] averages the pooling features by normalization and concatenation and form context vector to append spatial information on subsequent layers of the network. To capture the high-level global context and minimize the reduction of signal resolution, PixelNet [20] combines the ideas of [16] and [19]. PixelNet concatenates spatial context from multiple layers to build hypercolumn and randomly sample a small number of pixels instead of the whole image in training phase. However, PixelNet doesn't consider imbalance classes of the dataset while sampling pixels for training. In this work, we propose a class balance technique by sampling pixels equally for all the classes with PixelNet and utilize it for brain tumor and stroke lesion segmentation.

## 2. METHOD

PixelNet (fig. 1) follows VGG-16[15] architecture which forms with 13 convolutional layers and three fully-connected (fc) layers. In PixelNet, last two fc layers transform to convolutional filters as [21] and aggregate them as a set of convolutional features to form multi-scale hypercolumn descriptor. Finally, hypercolumn features feed to a multi-layer perception (MLP) with 3 fully connected layers of size 4096 followed by ReLU activations where K classes in the last layer output prediction. The hypercolumn descriptor can be written as-

$$h_p = [c_1(p), c_2(p), ..., c_M(p)]$$

where $h_p$ denotes the multiscale hypercolumn features for the pixel $p$, and $c_i(p)$ denotes the feature vector from layer $i$. As it considers pixel-wise prediction, so the final prediction for a pixel p, $f_{\theta,p}(X) = g(h_p(X))$, where both hypercolumn features $h$ and pixel-wise predictor $g$ have been represented by $\theta$ which update using SGD training. As a non-linear predictor, it adopts sparse pixel prediction at training time for efficient mini-batch generation. In sparse prediction, hypercolumn features $h_p$ choose from dense convolutional responses at all layers by computing the 4 discrete locations in the feature map $c_i$ (for $i^{th}$ layer) closest to sampled pixel $p \in P$ and finally apply bilinear interpolation to get $i^{th}$ layer response in hypercolumn.

### 2.1 Class-Balanced PixelNet

Most of the medical data contain unequal number of class labels in training data which consider as imbalance dataset. The main disadvantage of unbalance training data in convolutional learning is that it often leads to converge in certain labels. PixelNet samples N (variable) pixels per image randomly without considering consisting of pixels from each class in training phase. In our approach, it ensures all existing class labels inside N pixels while sampling for training from each image. If there are K classes in an input image, then we sample N/K pixels for every class to ensure an equal number of pixels to train the model. However, there are some samples where class skewness is very high. In this case, we sample randomly to maintain the equal number of sampling pixel.

## 3. EXPERIMENTS

### 3.1 Dataset

As most of the diagnostics of brain malformation use magnetic resonance (MR) scan, so we utilize our approach on two public brain MRI datasets such as BraTS 2017 (Brain Tumor Image Segmentation Benchmark) [22, 23, 24, 25] and ISLES 2017 (Ischemic Stroke Lesion Segmentation).

BraTS 2017 consists 285 training and 46 validation MRI scan of both high-grade glioma (HGG) and low-grade glioma (LGG). It has 4 different modalities including T1 (spin- lattice relaxation), T1c (T1-contrasted), T2 (spin-spin relaxation) and FLAIR (fluid attenuation inversion recovery) and each scan contains 3D volume of size 155x240x240 (155 slices). The ground-truth with manual segmentation contains three classes like GD-enhancing tumor (ET — label 4), the peritumoral edema (ED — label 2), and the necrotic and non-enhancing tumor (NCR/NET — label 1) where background (label 0) contains healthy tissue and padding.

ISLES 2017 consists of 43 training and 32 testing MRI scans of Ischemic stroke patients. It obtains 7 different modalities of diffusion and perfusion maps which includes DWI (Diffusion weighted imaging), ADC (Apparent diffusion coefficient), CBV (Cerebral blood volume), CBF (Cerebral blood flow), MTT (Mean transit time), TTP (Time to peak) and TMax (Time to maximum). The ground-truth contains only lesion part (label 1) where background (label 2) is healthy brain tissue and padding. It consists variable voxel size such as 19 to 30 2D slices of size 128x128 to 256x256. However, both BraTS and ISLES are highly imbalance where around 98% pixels are background or healthy tissues.

## 3.2 Preprocessing

There are different kinds of MRI artifacts that affect the performance of tumor segmentation algorithm. Most often artifact is bias field distortion which leads an inhomogeneous background in MR images. We apply N4ITK [26] filter to remove the bias field distortion. The data is then normalized within each input channel by subtracting the channel's mean and dividing by the channel's standard deviation to have zero mean and unit variance.

## 3.3 Training and Evaluation

We use depth slicing 2D images on axial orientation of MR images and corresponding manually segmented ground truth to train our model. To initialize model parameters, we use VGG-16 [15] pre-trained model. We use Caffe deep learning platform to perform all our experiments and evaluate the performance score by using online portal of corresponding challenges.

### 3.3.1 Brain Tumor Segmentation (BraTS)

As BraTS consists 3 modalities of MRI sequences like Flair, T1c and T2, so we utilize 3 channels of input, one slice from each. During training phase, multi-scale convolutional features of N (2000) sample pixels have been extracted to form hypercolumn. To cope with class imbalance problem, we sample N with equal number of pixels for each class (in this case, maximum K = 4 classes) consists of the corresponding slice. In testing phase, entire image feed-forwards to predict segmentation from last layer. Finally, predicted segmentation evaluate by the online portal of BraTS 2017 challenge.

**Table 1. Evaluation score for BraTS 2017 validation data**

|  | Enhance(ET) | Whole(WT) | Core(TC) |
| --- | --- | --- | --- |
| Dice | 0.701 | **0.879** | 0.781 |
| Sensitivity | 0.730 | 0.865 | 0.721 |
| Specificity | 0.999 | 0.997 | 0.998 |
| Hausdorff95 | 11.921 | 9.110 | 11.381 |

Evaluation scores of validation data have shown in the table 1 where are quite competitive results in BraTS 2017 challenge. Table 2 represents the comparison of our class balanced PixelNet and original PixelNet. It is clear that class balanced PixelNet outperforms original PixelNet in all the categories.

**Table 2. Dice and Hausdorff95 of class balanced PixelNet and original PixelNet with BRATS 2017 validation dataset**

|  | Dice | | | Hausdorff95 | | |
| --- | --- | --- | --- | --- | --- | --- |
|  | ET | WT | TC | ET | WT | TC |
| Class Balanced PixelNet | **0.701** | **0.879** | **0.781** | **11.921** | **9.110** | **11.381** |
| PixelNet | 0.689 | 0.876 | 0.761 | 12.938 | 9.820 | 12.361 |

### 3.3.2 Stroke Lesion Segmentation (ISLES)

We utilize only 3 modalities like DWI, ADC, and CBF among 7 provided modalities in ISLES 2017 dataset. In training, we use 2D axial slices by upsampling to 256x256 with zero padding. As BraTS, we use N (2000) sample pixels to form hypercolumn. We sample N with equal number of classes K (in this case K=2) to handle class imbalance problem. Later entire image uses as an input to predict stroke lesion segmentation in testing phase. We evaluate the segmentation performance score by using online portal of the ISLES 2017 challenge.

**Table 3. Evaluation score for ISLES 2017 testing data**

| Dice | Hausdorff | Avg. Dist. | Precision | Recall |
| --- | --- | --- | --- | --- |
| **0.29 ±0.2** | 46.9 ±12.8 | 7.3 ±3.02 | 0.35 ±0.23 | 0.6 ±0.24 |

Table 3 represents the evaluation results of testing data where class balanced PixelNet denotes as a competitive method in ISLES 2017 challenge. The reasons behind the poor performance of testing data are small dataset, small lesion, and less clinical information. So, lack of dataset leads the model towards overfitting which is main reason for overall poor performance of ISLES 2017 challenge. Besides, absolute intensity changes in different subjects. A comparison between our model and origin PixelNet has been demonstrated in table 4. Class balanced PixelNet shows better performance in both dice and hausdorff measurement. Fig. 2 shows predicted segmentations for both datasets. From this figure, we can see that Class Balanced PixelNet performs state of art accuracy in the brain tumor segmentation and shows average performance for stroke lesion segmentation considering improper dataset.

**Table 4. Evaluation score for ISLES 2017 testing data**

| Model | Dice | Hausdorff | Avg. Dist. | Precision | Recall |
| --- | --- | --- | --- | --- | --- |
| Class Balanced PixelNet | **0.29** | **46.92** | **7.31** | **0.35** | **0.62** |
| PixelNet | 0.28 | 76.6 | 11.35 | 0.30 | 0.58 |

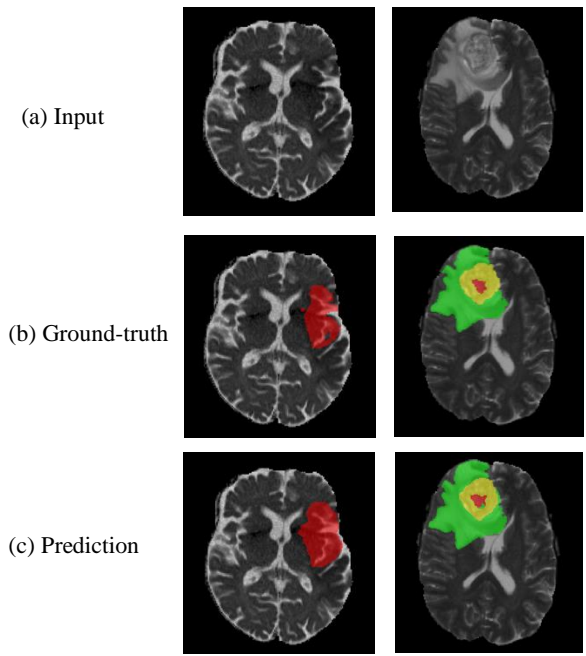

**Figure 2. Example of Class balanced PixelNet prediction on training dataset. Right (a, b, c) Brain tumor segmentation where predict necrotic (red), enhancing(yellow) and edema (green). Left (a, b, c) stroke lesion segmentation where predict stroke lesion area (red).**

## 4. CONCLUSION

We present an automatic brain MR image segmentation model based on pixel-level prediction using small number of pixels rather than whole image or patch. We utilize our model in two brain MRI datasets like BraTS and ISLES and able to achieve promising. As PixelNet use small number of pixels during training time, so we can perform more experiments by utilizing smaller dataset in future. Though class balanced PixelNet shows state of art performance for the whole tumor segmentation, it is still demanding to increase the accuracy of other regions like enhance tumor (ET) and tumor core (TC). As MRI is a 3D imaging technique, so we are planning to apply 3D PixelNet for future work with adopting class balance technique. Moreover, MRI scans consist multi-modalities image for example 4 modalities for BraTS and 7 modalities for ISLES, where current deep learning technique can achieve maximum performance with only 3 modalities and accuracy almost similar even utilizing more modalities. We are also planning to utilize multi-modalities efficiently so that it can achieve human-level accuracy for medical imaging.

## 5. ACKNOWLEDGMENTS

This work is supported by the Singapore Academic Research Fund under Grant R-397-000-227-112, NUSRI China Jiangsu Provincial Grant BK20150386 & BE2016077 and NMRC Bedside & Bench under grant R-397-000-245-511 awarded to Dr. Hongliang Ren. I would like to thank Aayush Bansal, the author of PixelNet [25], for the assistance to implement PixelNet for this project.